\documentclass[11pt,reqno]{amsart}
\usepackage{amsmath,amssymb,amsthm,amscd, amsfonts, mathrsfs}

\usepackage{cases}

\usepackage[dvips]{epsfig}
\usepackage{epsf}
\usepackage{verbatim} 

\usepackage[bookmarksnumbered,pdfpagelabels=true,plainpages=false,colorlinks=true,
            linkcolor=black,citecolor=black,urlcolor=blue]{hyperref}

\theoremstyle{plain}
\newtheorem{theorem}{Theorem}[section]

\theoremstyle{definition}
\newtheorem{remark}[theorem]{Remark}

\newtheorem{definition}[theorem]{Definition}

\numberwithin{equation}{section}

\newcommand{\cF}{\mathcal F}

\newcommand{\cI}{\mathcal I}

\newcommand{\cL}{\mathcal L}

\newcommand{\cS}{\mathcal S}

\newcommand{\cU}{\mathcal U}

\newcommand{\ccP}{\mathscr{P}}

\newcommand{\al}{\alpha}
\newcommand{\be}{\beta}
\newcommand{\ga}{\gamma}
\newcommand{\Ga}{\Gamma}
\newcommand{\de}{\delta}

\newcommand{\la}{\lambda}

\newcommand{\si}{\sigma}
\newcommand{\Si}{\Sigma}

\newcommand{\Om}{\Omega}

\newcommand{\RR}{\mathbb R}

\newcommand{\rar}{\rightarrow}

\newcommand{\p}{\parallel}

\newcommand{\K}{\mathscr{K}}

\newcommand{\mss}{\hspace{0.2cm}}

\newcommand{\bv}{\kappa}
\newcommand{\sv}{\vartheta}
\newcommand{\bm}{\mathbf{m}}
\newcommand{\bs}{\mathbf{s}}
\newcommand{\bt}{\mathbf{t}}

\allowdisplaybreaks

\title[Relativistic viscous fluids]{On the well-posedness of relativistic viscous fluids}

\author[Disconzi]{Marcelo M. Disconzi}
\address{Department of Mathematics\\
Vanderbilt University, Nashville, TN 37240, USA}
\email{marcelo.disconzi@vanderbilt.edu}
\thanks{The author is supported by NSF grant 1305705.}

\begin{document}

\maketitle

\begin{abstract}
Using a simple and well-motivated modification of the stress-energy tensor
for a viscous fluid proposed by Lichnerowicz, we prove
that Einstein's equations coupled to a relativistic version of the
Navier-Stokes equations are well-posed in a suitable Gevrey class 
if the fluid is incompressible and irrotational. These last two conditions are
given an appropriate relativistic interpretation. 
The solutions enjoy the domain of dependence or finite propagation
speed property.
We also derive a full set of equations,
describing a relativistic fluid that is not necessarily incompressible or irrotational,
which is well-suited for comparisons with the system of an inviscid fluid. 
\end{abstract}

\section{Introduction. \label{intro}}

It has been known for a long time that 
one cannot account for some important features of 
cosmological and astrophysical phenomena without 
 incorporating dissipation into our physical models 
 \cite{Lich_book_GR,MTW,Weinberg_GR}.
More recently, advances in the study of heavily dense 
objects, such as neutron stars \cite{CSS,DLSS,SG}, and in our understanding
of the dynamics of the early universe \cite{BB,Ma}, 
point toward the necessity of a relativistic
description of dissipative phenomena. Experience
with General Relativity suggests that the correct approach 
to this question should rely on the construction of a stress-energy 
tensor which subsumes characteristics due to the viscosity of the medium
under consideration.

In spite of that, we still lack a satisfactory 
formulation of viscous phenomena within Einstein's 
theory of General Relativity. One of the main reasons for this is 
the lack of a variational formulation of 
the classical (non-relativistic) Navier-Stokes equations.
In the absence of a variational description of the equations of motion,
one does not have a principle that uniquely defines the stress-energy
tensor $T_{\al\be}$ in the context of General Relativity. As a consequence,
there have been different proposals for what the correct  
$T_{\al\be}$ should be. We refer the reader to \cite{Ma} and references
therein for 
a brief history of different attempts to formulate a viscous relativistic
theory. A more complete and up-to-date discussion can be found in \cite{RZ}.

Still, 
in many respects, the natural choice for a viscous stress-energy
tensor seems to be
\begin{gather}
T^N_{\al\be} = (p + \varrho) u_\al u_\be - p g_{\al\be} +
\bv \pi_{\al\be} \nabla_\mu u^\mu + \sv \pi_\al^\rho \pi_\be^\mu 
(\nabla_\rho u_\mu + \nabla_\mu u_\rho),
\label{T_natural}
\end{gather}
where
$p$ and $\varrho$ are respectively the
pressure and density of the fluid, $u$ is its four-velocity,
the bulk viscosity $\bv$ and the shear viscosity  $\sv$ are 
non-negative constants, 
$g$ is a Lorentzian metric\footnote{Our convention for the metric is $(+---)$.}
  and 
 $\pi_{\al\be} = g_{\al\be} - u_\al u_\be$.
$p$ and $\varrho$ are related by an equation known as equation of state, 
the choice of which depends on the nature of the fluid and
 has to be given in order to close the system of
the equations of motion (see below).
We say that the choice $T^N_{\al\be}$ is natural because it is a straightforward 
covariant generalization of the stress-energy tensor of a viscous
non-relativistic fluid\footnote{We remind the reader that the stress-energy
tensor for a non-relativistic viscous fluid is known, despite the absence
of a variational formulation of the classical Navier-Stokes equations. It 
is  constructed by exploring the conservation of mass, energy, and 
momentum of the problem. A similar procedure becomes
ambiguous in the setting of General Relativity.}, and it reduces to the stress energy tensor 
of an inviscid fluid\footnote{Which is derived from a variational
approach. We remark that since no other interactions will be added, 
we use the terms inviscid and ideal as synonymous.} when $\bv = \sv = 0$.

Unfortunately, as Pichon demonstrated \cite{Pichon}, 
the equations of motion derived from Einstein's equation
coupled to (\ref{T_natural}) exhibit superluminal  signals
when $p + \varrho \gg 1$, a feature unacceptable for a relativistic theory.
Other attempts to formulate a viscous relativistic theory based on 
a simple covariant generalization of the classical (i.e., non-relativistic)
stress-energy tensor for the Navier-Stokes equations
 have also
failed to produce a causal theory. See \cite{RZ} for a detailed discussion.

One way of overcoming the lack of causality in such models consists 
in extending  the variables of the theory. This leads to what
is known as Relativistic Extended Irreversible Thermodynamics.
In these approaches, it is possible to show that, under certain
 circumstances,  the 
equations of motion fall into 
known classes of hyperbolic equations, exhibiting, as consequence, finite propagation speed.
It is not at all clear, however, that the equations remain hyperbolic under all
physically realistic scenarios. Furthermore, the plethora of models that comes out
of the extended thermodynamic approach suggests that it entails many
\emph{ad-hoc} features, in sharp contrast to the usually unique way of coupling
gravity to matter via the introduction of the stress-energy tensor of matter 
fields (when the latter
 is uniquely determined by a variational characterization).
For the sake of this discussion, we shall 
refer to the procedure of coupling gravity to matter
by solely introducing the stress-energy tensor of matter fields 
on the right hand side of Einstein's equations,
without extending the  space of variables as in  
Relativistic Extended Irreversible Thermodynamics, as the
\emph{traditional} approach to the coupling of gravity and matter.
We refer the reader to \cite{JCL,MR,RZ} for a treatment of
Relativistic Extended Irreversible Thermodynamics.

It is also important to mention the rather curious
fact that some of the aforementioned recent advances in the 
modeling of heavy objects were accomplished by numerically solving Einstein's 
equations coupled to essentially (\ref{T_natural}), and working in a regime where 
the superluminal problem could be avoided \cite{DLSS}. The ability to generate 
physically relevant results in a context where it is expected that viscosity
will play a crucial role, indicates that the traditional 
approach to the problem
should not be dismissed, despite all failed attempts in producing a causal theory
when viscosity is present.

It is, therefore, worthwhile to take a fresh look at the question
of whether there is a correct stress-energy tensor $T_{\al\be}$ that describes relativistic viscous
fluids, and that can be coupled to gravity in the traditional way.
 The above considerations suggest that if there is a correct choice
for $T_{\al\be}$, it should
be close to $T^N_{\al\be}$ for appropriate values of the quantities involved.
``Appropriate," here, means choices of these quantities
that lead to good agreement between models derived from 
(\ref{T_natural}) and observational data.

We propose the following guiding 
principles
in the search for $T_{\al\be}$, namely, that any
candidate for a stress-energy tensor of a relativistic viscous fluid 
should satisfy:
(i) it reduces to $T_{\al\be} = 
(p + \varrho ) u_\al u_\be - p g_{\al\be}$, i.e., to the stress-energy tensor
of an ideal fluid,  when dissipation is absent; (ii) it reduces to the
stress-energy tensor of the classical Navier-Stokes equations in the non-relativistic
limit; (iii) it is close
to $T^N_{\al\be}$ in the sense mentioned above;
(iv) the equations of motion derived from coupling Einstein's equations
to $T_{\al\be}$ are well-posed and exhibit the domain of dependence
property, with the speed of propagation of disturbances 
being at most the speed of light\footnote{A different approach to the 
construction of the equations of motion for a relativistic viscous
fluid has been adopted in \cite{Geroch}. Notice that this had been proposed 
prior to the developments cited here, specially those dealing with the numerical 
investigation of the problem. It is not entirely clear to this author
how compatible the approach taken in \cite{Geroch} and these numerical results 
are.}.

Consider the following stress-energy tensor:
\begin{align}
T_{\al\be} & = (p + \varrho ) u_\al u_\be - p g_{\al\be}
+ \bv \pi_{\al\be} \nabla_\mu C^\mu 
+ \sv\pi_\al^\rho \pi_\be^\mu (\nabla_\rho C_\mu + \nabla_\mu C_\rho).
\label{T_vis}
\end{align}
One immediately sees that (\ref{T_vis}) resembles (\ref{T_natural}),
with the terms in derivatives of  $u$ replaced by 
(derivatives of) what is known as the dynamic velocity $C$
($C$ is also called the current of the fluid;
$\pi_{\al\be}$ is still given by $\pi_{\al\be} = g_{\al\be} -
u_\al u_\be$), defined by
$C_\al = F u_\al$, where $F$ is the so-called index of the fluid.
It is defined as follows. It is customary to 
introduce\footnote{We shall not review the several thermodynamic aspects
of relativistic fluids necessary for this work. The reader can consult, for 
instance, \cite{FriRen,RZ}.}
the rest mass density $r$ and the specific internal energy
$\epsilon$ by $\varrho = r(1+\epsilon)$. Then $F = 1 + \epsilon + \frac{p}{r}$, 
so that $\varrho + p = r(1+\epsilon + \frac{p}{r}) = r F$.
$F$ and $C$ have been introduced by
Lichnerowicz in his proof of well-posedness of the equations
of relativistic hydrodynamics and 
magneto-hydrodynamics\footnote{Some interesting geometric 
consequences also follow from the introduction of the dynamic velocity;
see the above references.} \cite{Lich_fluid_1,Lich_fluid_2,Lich_MHD}.
For $F$ near one, (\ref{T_vis}) is close to (\ref{T_natural});
this will be the case, for example, for a large class of barotropic fluids,
 or for a 
fluid with small specific enthalpy, defined 
as $\epsilon + \frac{p}{r}$. 
From these remarks it follows that the stress-energy
tensor given by (\ref{T_vis}) is in line with the principles
(i)-(iii) stated above. In order to address the more delicate point
(iv), we need several further definitions and hypotheses.

A fluid with stress-energy tensor (\ref{T_vis}) will be called
incompressible if $\nabla_\mu C^\mu = 0$, in which case $T_{\al\be}$ 
becomes
\begin{align}
T_{\al\be} & = (p + \varrho ) u_\al u_\be - p g_{\al\be}
+ \sv\pi_\al^\rho \pi_\be^\mu (\nabla_\rho C_\mu + \nabla_\mu C_\rho).
\label{T_vis_i}
\end{align}
This definition of incompressibility is made so that
it agrees with the notion of an incompressible fluid when $\bv = \sv = 0$:
a relativistic inviscid barotropic fluid is said to be incompressible if its acoustic 
waves propagate at the speed of light. It is possible to show that this
is equivalent to the vanishing of the divergence of $C$.

We define the vorticity tensor by
\begin{gather}
\Om_{\al\be } = \nabla_\al C_\be - \nabla_\be C_\al 
\equiv \partial_\al C_\be - \partial_\be C_\al.
\nonumber
\end{gather}
A fluid is called irrotational if $\Om = 0$. For an incompressible
irrotational fluid, (\ref{T_vis}) simplifies to 
\begin{align}
T_{\al\be} & = (p + \varrho ) u_\al u_\be - p g_{\al\be} 
+ 2 \sv\pi_\al^\rho \pi_\be^\mu \nabla_\rho C_\mu.
\label{T_vis_ii}
\end{align}
Finally, we need to be more specific about the thermodynamic quantities 
$p$, $\varrho$, $r$, $\epsilon$, $F$, and the relations among them.
We suppose the validity of the the first law of thermodynamics,
in which case further thermodynamic variables, namely, 
the specific entropy $s$ and the absolute temperature $\theta$ are introduced.
In order to be consistent with guiding principle (i) above, we 
follow 
the standard approach of assuming that only two
of the thermodynamic quantities are independent, with the other ones
determined by thermodynamic relations among them coming
from the first law of thermodynamics and an equation of state, which
depends on the nature of the fluid. On physical grounds, all such relations
should be invertible, which renders the question of which two quantities 
are independent a matter of choice. Here we shall assume that 
$r$ and $s$ are independent and postulate an equation of state 
of the form
\begin{gather}
\varrho = \ccP(r,s).
\label{eq_state_r_s}
\end{gather}
It follows that $p = p(r,s)$, 
$\theta=\theta(r,s)$, $\epsilon=\epsilon(r,s)$, 
and $F=F(r,s)$ are known if $r$ and $s$ are. We notice,
however, that later on it will be more convenient to treat $s$ and $F$
as independent variables, in which case we shall assume that the equation
of state takes the form $r = r(F,s)$. 
For physically relevant equations of state $F > 0$,
which allows us to restrict to positive values when treating $F$ as
an independent variable.
In this situation, the following
condition will be assumed to hold:
\begin{gather}
\frac{\partial r}{\partial F} \geq  \frac{r}{F},
\label{sound_speed_condition}
\end{gather} 
in particular $\frac{\partial r}{\partial F} > 0$ if
$r > 0$. 
Condition (\ref{sound_speed_condition}) 
has to be satisfied if we want to recover the stress-energy
tensor of an ideal fluid when $\bv = \sv = 0$, in that
it expresses the condition that sound waves in an ideal fluid
travel at most at the speed of light.
At last, we suppose that the equation of state is such that the temperature
satisfies
\begin{align}
\begin{split}
\theta(r,s) > 0 \text{ if } r > 0, s \geq 0, \\
\theta(F,s) > 0 \text{ if } s \geq 0, F > 0,
\end{split}
\label{temperature}
\end{align}
expressing the positivity of the temperature regardless of  the choice
of independent variables.

The system of equations to be studied consists of Einstein's equations coupled to 
(\ref{T_vis}), or (\ref{T_vis_i}), or (\ref{T_vis_ii}), and reads
\begin{subnumcases}{\label{original}}
R_{\al\be} - \frac{1}{2}R g_{\al\be} = \K T_{\al\be},
\label{original_Einstein} \\
\nabla^\al T_{\al\be} = 0,
\label{original_conservation} \\
\nabla_\al( r u^\al )  = 0, 
\label{original_mass_conservation} \\
u^\al u_\al = 1.
\label{original_normalization}
\end{subnumcases}
$R_{\al\be}$ and $R$ are, of course, the Ricci and scalar curvature
of the metric $g$, and $\K$ is a constant. 
Equation (\ref{original_normalization}) is the standard
normalization condition on the velocity of a relativistic fluid,
whereas (\ref{original_mass_conservation}) expresses
the condition that mass is locally conserved along the flow lines. 
We notice that without introducing
(\ref{original_mass_conservation}), the motion of the fluid is underdetermined.
The unknowns are the metric $g$, the fluid velocity $u$, which is a vector field,
the specific entropy $s$, and the rest mass density $r$. These last two quantities 
are  non-negative
real valued functions. We suppose that we are also given a
smooth function $\ccP: \RR_+ \times \RR_+ \rar \RR$ that gives the 
 equation of state (\ref{eq_state_r_s}), with all the other thermodynamic quantities 
introduced above given as functions of $s$ and $r$ via relations derived 
from the first law of thermodynamics and the equation of state.

\begin{definition}
System (\ref{original}) with $T_{\al\be}$ 
given by (\ref{T_vis}) will be called the Einstein-Navier-Stokes
system; incompressible Einstein-Navier-Stokes when $T_{\al\be}$
is given by (\ref{T_vis_i}); and  
incompressible irrotational Einstein-Navier-Stokes system when $T_{\al\be}$
assume the form  (\ref{T_vis_ii}).
\end{definition}

\noindent \textbf{Assumption.} 
We shall assume for the rest of the text that $\sv > 0$. \\

An initial data set for the Einstein-Navier-Stokes system consists of 
a three-dimensional  manifold $\Si$, a Riemannian metric\footnote{Except that
with our conventions this metric is negative definite.}
 $g_0$, a symmetric two-tensor $\kappa$, two real valued non-negative
 functions $s_0$ and $r_0$, and a vector field $v$; these are all quantities 
 defined on $\Si$. As it is well-known, these data cannot be arbitrary
 but must satisfy the constraint equations, which read
 in a coordinate system with $\partial_0$ transversal and
 $\partial_i$, $i=1,2,3$, tangent to $\Si$, as
$S_{\al 0 } = \K T_{\al 0 }$, where 
$S_{\al\be} = R_{\al\be} - \frac{1}{2}R g_{\al\be}$
is the Einstein tensor. By definition, an initial data set
always satisfies these constraints.

We are now ready to state the main result. We refer the 
reader to the standard literature in General Relativity for the terminology
employed in Theorem \ref{main_theorem}, and to appendix \ref{Leray_Ohya} or references
\cite{Leray_Ohya,Ro} for the definition of the Gevrey spaces 
$\ga^{m, (\si)}$.

\begin{theorem}
Let $\cI = (\Si, g_0, \kappa,v, s_0,  r_0)$ be an initial data set for the 
 incompressible irrotational Einstein-Navier-Stokes system, 
 with $\Si$ compact,  $s_0 > 0$, $r_0 > 0$, and an equation of state $\ccP$
such that (\ref{sound_speed_condition}) and (\ref{temperature}) 
are satisfied. 
Assume that the initial data is in 
$\ga^{(\si)}(\Si)$ for some $1 \leq \si <2$.
Then there exist a space-time $(M,g)$
that is a development of $\cI$,
 real valued  functions $s > 0$ 
and $r > 0$
defined on $M$, and a vector field $u$, 
such that  
$g \in \ga^{9, (\si)}(M)$,
$u \in \ga^{8, (\si)}(M)$, 
$s \in \ga^{8, (\si)}(M)$, 
$r \in \ga^{8, (\si)}(M)$, and  
$(g,u, s, r)$ satisfy the  incompressible irrotational
Einstein-Navier-Stokes system  in $M$.

Furthermore, this solution satisfies the geometric uniqueness
and domain of dependence properties, in the following sense. 
Let $\cI^\prime = (\Si^\prime, g_0^\prime, \kappa^\prime, v^\prime, s_0^\prime,  r_0^\prime)$
be another initial data set, also with equation of state $\ccP$,
with corresponding development $(M^\prime,g^\prime)$ and 
solution $(g^\prime,u^\prime, s^\prime, r^\prime)$ 
of  the incompressible irrotational Einstein-Navier-Stokes equations
in $M^\prime$.
Assume  that there exists a diffeomorphism
 between $S \subset \Si$ and $S^\prime \subset \Si^\prime$ that carries
$\left. \cI\right|_S$ onto $\left. \cI^\prime \right|_{S^\prime}$, where $S$ and $S^\prime$
are, respectively, domains in $\Si$ and $\Si^\prime$. Then there exists
a diffeomorphism between $D_g(S) \subset M$
and  $D_{g^\prime}(S^\prime) \subset M^\prime$ carrying
$(g,u, s, r)$ onto
$(g^\prime,u^\prime, s^\prime, r^\prime)$, where $D_g(S)$ denotes
the future domain of dependence of $S$ in the metric $g$; in particular $D_g(S)$ and 
$D_{g^\prime}(S^\prime)$ are isometric.
\label{main_theorem}
\end{theorem}
We have chosen to work in the Gevrey class because the equations
we shall derive form a Leray-Ohya system\footnote{Called 
``hyperbolique non-stricts" by Leray and Ohya. Sometimes
these systems are called weakly or degenerate hyperbolic, although certainly
these terms have been used to denote different types of systems in the literature.} 
(see appendix \ref{Leray_Ohya}), which, in general, are not well-posed 
in Sobolev spaces.
The space-time $M$ is diffeomorphic to $\Si \times [0,T]$ for some $T > 0$, 
and to $\Si \times [0,\widetilde{T})$ for some $\widetilde{T} > T$
 if we require it to be a maximal Cauchy 
development. In light of the domain of dependence property, the compactness of $\Si$
is not absolutely necessary, although in the case of a 
non-compact $\Si$ without asymptotic conditions on the initial data, $M$ may not contain any Cauchy surface other than 
$\Si$ itself. The hypotheses $s_0>0$ and $r_0 > 0$ guarantee, by continuity, 
the positivity of $s$ and $r$ in the neighborhood of $\Si$, as stated in the theorem.
The assumption $s_0 > 0$ could be weakened to $s_0 \geq 0$, but in this case 
the non-negativity of $s$ in $M$ would have to be derived from the equations of
motion, a task we avoid for brevity. Allowing $r_0$ to vanish, however, causes severe
difficulties, and the well-posedness of the corresponding problem is largely 
open even in the case of an ideal fluid.

The stress-energy tensor (\ref{T_vis}) was first introduced
by Lichnerowicz \cite{Lich_book_GR}, except that it contained an extra term of the form
$\sv \pi_{\al\be} u^\mu \partial_\mu F$. As it was pointed out by
Lichnerowicz himself and later by Pichon \cite{Pichon}, this extra term leads to an 
 indetermination in the computation of the pressure. Pichon proposed
 subtracting this term, leading in this way to (\ref{T_vis}).
 The reader is referred to their original works for the physical insights leading
 to the construction of (\ref{T_vis}). Choquet-Bruhat has also proposed 
a stress-energy tensor similar to (\ref{T_vis}) \cite{CB}. Her proposal, however,
does not include the projection terms $\pi_{\al\be}$, and the viscous
terms are, therefore, \emph{linear} in the velocity.

We finish this introduction with some comments on the hypotheses and the thesis
of Theorem \ref{main_theorem}. Perhaps the first hypothesis one would like
to remove is that of an irrotational fluid. 
While this assumption is 
certainly unsatisfactory from a physical point of view, 
we remind the reader that we are attempting, in a sense, to ``reboot" the 
traditional approach to the problem. In other words, we try to identify 
a suitable candidate for $T_{\al\be}$ that leads to a causal theory, without
relying on the introduction of extra variables, as in the extended irreversible 
thermodynamic models. It is quite natural, therefore, to start by analyzing
a simpler situation\footnote{Although, as the reader can check below, already under the 
present assumptions, the system of equations is rather involved.}.
Another hypotheses we would like to weaken is the use of initial data 
in Gevrey spaces\footnote{It will be shown in a future work that, upon restricting 
to a smaller Gevrey class than as in Theorem \ref{main_theorem},
the irrotational hypothesis can be removed \cite{CD_prep}.}. 
These spaces have become an important tool in analyzing the 
equations of Fluid Dynamics, specially when viscosity is present 
(see, e.g., \cite{BBT,CRT,FT,Ro} and references therein). Hence, it is sensible
that such spaces might play a role in the case of relativistic viscous fluids as well. 
Furthermore, Gevrey spaces are not completely foreign to the study of Einstein's equations:
in some relevant circumstances, the equations of ideal magneto-hydrodynamics appear to 
have been shown to be well-posed only in the 
Gevrey class \cite{CB,FriRen}\footnote{Although it is very likely that
the formulation of \cite{AP} would carry over, with almost no modifications, to the coupling
with Einstein's equations. A proof of this statement, however, does not
seem to be available in the literature.}. On the other hand,
the overwhelming success of Sobolev space techniques in the investigation of the Cauchy problem
for Einstein's equations\footnote{The literature on this topic is too vast; see, e.g., the monographs 
\cite{CB,Ri}.} almost demands that we employ Sobolev spaces in the study of the evolution
problem. Moreover, in order to eventually settle the question of 
whether (\ref{T_vis}) can give a physically satisfactory description
of relativistic viscous phenomena, we have to be able to explicitly compute several physical
observables. For this, one has to solve the equations numerically, which, in turn,
requires that the equations be well-posed in some function space characterized by 
a finite number of derivatives.

Such restrictions notwithstanding, one should not overlook the 
conclusion of Theorem \ref{main_theorem}: it is possible, employing what 
we called a traditional  approach  (i.e., one 
the avoids the introduction of extra physical variables as 
in Relativistic Extended Irreversible Thermodynamics), to obtain 
a description of relativistic viscous fluids that satisfies 
the natural requirements (i)-(iv) discussed above. In particular,
the equations of motion are well-posed, and they do not 
exhibit faster than light signals.

This paper is organized as follows. In section \ref{new_system_section} we derive
from (\ref{original}) new systems of equations for both the incompressible
and incompressible irrotational system. Initial data for these 
systems are calculated from the original initial data set 
in section \ref{initial_data_section}. The characteristics of both systems
are also studied in this section. In section \ref{proof_secion}, 
we prove Theorem
\ref{main_theorem}. For this we shall use the results of Leray and Ohya 
\cite{Leray_Ohya} that are
reviewed in appendix \ref{Leray_Ohya} for the reader's convenience.
Finally, in section \ref{comparison_section}, we derive yet another system
of equations that is suited for comparisons with the case of an inviscid fluid.
We remark that although Theorem \ref{main_theorem} applies
only to the incompressible irrotational Einstein-Navier-Stokes system,
many of the arguments below (e.g., the derivation of the initial data) are carried
out in the more general case of a fluid that is only incompressible, with the 
condition of zero vorticity introduced at a later point. Along the way we shall obtain that
at least for analytic Cauchy data, the incompressible (not necessarily irrotational)
Einstein-Navier-Stokes system can be solved.

In the following, we adopt:

\noindent \textbf{Convention.} Greek indices run from $0$ to $3$ and Latin indices
from $1$ to $3$.

\section{A new system of equations.\label{new_system_section}}

In this section we assume that we are given a solution
to the Einstein-Navier-Stokes system. We  suppose 
that $s>0$ and $F>0$ are the two independent thermodynamic variables,
 that the equation of state reads $r =r(F, s)$
and satisfies the hypotheses stated in Theorem \ref{main_theorem}.

Using thermodynamic relations to express $\varrho$ and $p$ in terms of 
$s$ and $F$ and equation  (\ref{original_mass_conservation}),
it is seen that (\ref{original_conservation}) 
  decomposes as
\begin{gather}
\frac{r \theta}{F} C^\al \partial_\al s + \bv L^{(s)} + \sv  V^{(s)} = 0,
\label{entropy_eq}
\end{gather}
and 
\begin{gather}
\frac{r}{F} C^\al \Om_{\al\be} + \theta r \partial_\be s - \frac{r \theta}{F^2}
C_\be C^\al \partial_\al s + \bv L_\be + \sv V_\be = 0,
\label{contraction_Omega}
\end{gather} 
where $\theta = \theta(F,s)$ is the temperature,
\begin{gather}
L^{(s)} = F^{-2}(\frac{1}{F} C^\tau C^\mu \nabla_\tau C_\mu - F \nabla_\mu C^\mu)
\nabla_\rho C^\rho,
\nonumber
\end{gather}
\begin{gather}
V^{(s)} = -\frac{1}{F} (g_\al^\mu - F^{-2} C_\al C^\mu) (\nabla_\mu C_\nu + 
\nabla_\nu C_\mu) C^\be \nabla^\al (F^{-2} C_\be C^\nu), 
\nonumber
\end{gather}
\begin{align}
\begin{split}
L_\be & = (g_{\be}^\al - F^{-2} C_\be C^\al) \nabla_\al \nabla_\mu C^\mu 
\\
& 
- F^{-2}(g_\be^\ga - F^{-2} C_\be C^\ga)
C^\al \nabla_\al C_\ga  \nabla_\mu C^\mu ,
\end{split}
\label{L_be}
\end{align}
\begin{align}
\begin{split}
V_\be & = -(g_\be^\mu - F^{-2} C_\be C^\mu)(\nabla_\rho C_\mu + \nabla_\mu C_\rho)
\nabla_\al(F^{-2} C^\rho C^\al)
\\ 
&  - (g^{\al\rho} - F^{-2} C^\rho C^\al)
(\nabla_\rho C_\mu + \nabla_\mu C_\rho)(g_\be^\ga - F^{-2} C_\be C^\ga)\nabla_\alpha(F^{-2}C^\mu C_\ga) 
\\
& +(g_\be^\mu - F^{-2}C_\be C^\mu)(g^{\al\rho} - F^{-2} C^\rho C^\al)
(\nabla_\al \nabla_\rho C_\mu + \nabla_\al \nabla_\mu C_\rho).
\end{split}
\label{V_be}
\end{align}
We use the superscript $^{(s)}$ to emphasize that $L^{(s)}$ and $V^{(s)}$
 scalars
that come from the same part of the stress-energy tensor as the terms
$L_\al$ and $V_\al$.
Equation (\ref{contraction_Omega}) can be written in invariant form as
\begin{gather}
\iota_C \Om + \theta F  ds - \frac{\theta }{F} C \iota_C ds + 
\bv \frac{F}{r}  L + \sv \frac{F}{r}  V = 0,
\nonumber
\end{gather}
were $\iota_C$ is the interior product with $C$.
Taking the exterior derivative and recalling
\begin{gather}
\cL_C \Om = (d \iota_C + \iota_C d ) \Om = d \iota_C \Om,
\label{flow_lines_vorticity}
\end{gather}
where  $\cL_X$ is the Lie derivative in the direction of the
vector field $X$ and we have used  $d\Om = 0$ ($\Om$ is a closed form by
the way it was defined), 
we obtain, after expressing the Lie derivative in terms of covariant
derivatives,
\begin{align}
\begin{split}
 C^\mu & \nabla_\mu \Om_{\al\be}   +
\nabla_\al C^\mu \Om_{\mu\be} + \nabla_\be C^\mu \Om_{\al\mu}
+ \partial_\al (\theta F ) \partial_\be s - \partial_\be (\theta F) \partial_\al s 
\\
& 
-C^\mu \partial_\mu s \Big ( \partial_\al \left(\frac{\theta}{F}\right)C_\be 
- \partial_\be \left( \frac{\theta}{F} \right) C_\al \Big )
-\frac{\theta}{F} C^\mu \partial_\mu s \Om_{\al\be} 
\\
& - \frac{\theta}{F}\Big(
( C^\mu \nabla_\mu \partial_\al s + \partial_\mu s \nabla_\al C^\mu ) C_\be 
- (C^\mu \nabla_\mu \partial_\be s + \partial_\mu s\nabla_\beta C^\mu )C_\al \Big)
\\
& 
+ \Big [ d \Big ( \bv \frac{F}{r} L + \sv \frac{F}{r} V \Big ) \Big ]_{\al\be} = 0.
\end{split}
\label{Helmholtz}
\end{align}
Equation (\ref{Helmholtz}) will be called relativistic viscous Helmholtz equation,
as it reduces to the standard Helmholtz equation for inviscid relativistic 
fluids when $\bv = \sv = 0$.

To complete the system, we shall consider the Laplacian of the current $C$.
Computing 
\begin{gather}
\Delta C = (d \de + \de d)C = d\de C+ \de \Om,
\label{Hodge_Laplacian}
\end{gather}
where $\de$ is co-differentiation. 
But
\begin{gather}
\Delta C_\al = -g^{\mu \rho}\nabla_\mu \nabla_\rho C_\al + R_\al^\mu C_\mu.
\label{Laplacian_curvature}
\end{gather}
Hence, using $(\de \Om)_\al  = - \nabla^\mu \Om_{\mu\al}$,
\begin{gather}
g^{\mu\rho} \nabla_\mu \nabla_\rho C_\al = \nabla^\mu \Om_{\mu\al}
+ \nabla_\al \nabla_\mu C^\mu + R_\al^\mu C_\mu.
\label{current_eq_original}
\end{gather}

Since $C^\mu C_\mu = F^2$, we can consider $F$ as a function of
$g_{\al\be}$ and $C^\al$. In this case, the system of equations to be studied consists
of the usual Einstein equations (\ref{original_Einstein}),
the entropy equation (\ref{entropy_eq}), the viscous relativistic Helmholtz 
equation (\ref{Helmholtz}), and the current equation (\ref{current_eq_original}).
These are 21 equations for the 21 variables: ten $g_{\al\be}$, one
$s$, six $\Om_{\al \be}$, and four $C_\al$.

In order to proceed further, 
we need to compute the term $\Big [ d \Big ( \bv \frac{F}{r} L + \sv \frac{F}{r} V \Big ) \Big ]_{\al\be} $.
From (\ref{V_be}), we see that this term is of
highest order in $\Om_{\al\be}$. Keeping all these terms renders
the system unpleasantly complex and difficult to analyze. We shall
make simplifying assumptions according to the statement of 
Theorem \ref{main_theorem}

\subsection{The system of  an incompressible fluid.\label{system_incompressible}}

We make our first simplifying assumption, namely, that the fluid is
incompressible:
\begin{gather}
\nabla_\mu C^\mu = 0.
\label{incompressible}
\end{gather}
It follows that $L^{(s)} = 0 = L_\al$. Also, $\nabla_\al \nabla_\mu C^\mu = 0$ 
and, therefore,
\begin{gather}
\nabla_\mu \nabla_\al C^\mu =R_{\al \mu} C^\mu .
\label{commutation}
\end{gather}
Equation (\ref{current_eq_original}) then reads
\begin{gather}
g^{\la\rho} \partial_{\la\rho} C_\al = B_\al(g, \partial g, \partial^2 g, 
\Om, \partial \Om, C, \partial C),
\label{current_eq_simple_second_order}
\end{gather}
where from now on we adopt the following:

\noindent \textbf{Convention.} The letters $B$ and $B^\prime$, with indices attached
when necessary, will be used to denote expressions where the 
number of derivatives of the variables $g$, $s$, $\Om$, and $C$ are
indicated in their arguments. The expression represented by the same
letter $B$ or $B^\prime$ can vary from equation to equation.

From (\ref{V_be}) we can then write
\begin{align}
\begin{split}
V_\al  & = (g_\al^\mu - F^{-2}C_\al  C^\mu)(g^{\si\rho} - F^{-2} C^\rho C^\si )
(\nabla_\si  \nabla_\rho C_\mu + \nabla_\si \nabla_\mu C_\rho) \\
& + B_\al(g, \partial g, C, \partial C).
\end{split}
\nonumber
\end{align}
Using (\ref{commutation}) produces, after some algebra,
\begin{align}
\begin{split}
V_\al & = g^{\la\si } \nabla_\la \nabla_\si C_\al  - F^{-2} C^\si C^\la \nabla_\la \nabla_\si 
C_\al 
\\
& + B_\al (g, \partial g, \partial^2 g, \partial^3 g, C, \partial C, \partial^2 C),
\end{split}
\label{V_high_order}
\end{align}
where the terms $\partial^2 g$ and $\partial^3 g$ are due to the presence
of $R_{\al \be}$ and its derivative, respectively.

Turning to the Helmholtz equation, we compute  $(d V)_{\al\be} = \partial_\al V_\be - \partial_\be V_\al$ using
(\ref{V_high_order}) and the definition of $\Om_{\al\be}$, obtaining
\begin{align}
\begin{split}
 g^{\la\si} 
\partial_{\la \si} \Om_{\al\be}  = 
& B^\prime_{\al\be} (g, \partial g, \partial^2 g, \partial^3 g, \partial^4 g, s,  \partial s,
\partial^2 s, 
\Om, \partial \Om,  C, \partial C, \partial^2 C, \partial^3 C).
\end{split}
\label{Helmholtz_simple}
\end{align}

Assume, as usual, that we wish to solve the reduced Einstein system,
which corresponds to Einstein's equations in harmonic coordinates, in which case (\ref{original_Einstein}) reads
\begin{gather}
g^{\mu\rho} \partial_{\mu \rho} g_{\al\be} = 
B_{\al\be}(g, \partial g, s, C, \partial C).
\label{Einstein_simple}
\end{gather}
Finally, the entropy equation (\ref{entropy_eq}) is of the form
\begin{gather}
C^\al \partial_\al s = B(g, \partial g, s, C, \partial C).
\label{entropy_eq_simple_first_order}
\end{gather}

Understanding, as before, that $F$ is a function of $C^\al$ and $g_{\al\be}$,
our system of equations for an incompressible fluid is given by  (\ref{current_eq_simple_second_order}), (\ref{Helmholtz_simple}), (\ref{Einstein_simple}),
(\ref{entropy_eq_simple_first_order}); the unknowns are $g_{\al\be}$, 
$s$, $\Om_{\al\be}$ and $C_\al$. The system, however, is not 
yet in a suitable form for an application of the Leray-Ohya
Theorem stated in the appendix. 
We shall take further derivatives of the equations and make one more 
simplifying hypothesis.
 
\subsection{The system of an  incompressible irrotational fluid.}
We consider now our second simplifying assumption, namely, that the fluid
is irrotational:
\begin{gather}
\Om = 0.
\label{irrotational}
\end{gather}
Applying $g^{\la\si}\partial_\la  \partial_\si$ to (\ref{Einstein_simple})
gives
\begin{gather}
g^{\la\si} g^{\mu\rho} \partial_{\la\si \mu \rho} g_{\al\be} = 
B_{\al\be}(g, \partial g, \partial^2 g, \partial^3 g, s, \partial s, \partial^2 s,
 C, \partial C, \partial^2 C).
\label{Einstein_simple_fourth_order}
\end{gather}
Notice that the right hand side does not contain third derivatives of $C$
because the fluid is incompressible and irrotational. In fact, 
these terms would come from
\begin{align}
\pi_\al^\rho \pi_\be^\mu & g^{\la\si} \partial_{\la\si} (\nabla_\rho C_\mu + \nabla_\mu 
C_\rho) = 
 \pi_\al^\rho \pi_\be^\mu g^{\la\si} \partial_{\la\si} (2 \nabla_\rho C_\mu 
+ \Om_{\mu \rho} ) \nonumber  \\
& = 
2  \pi_\al^\rho \pi_\be^\mu g^{\la\si} \partial_{\la\si}  \nabla_\rho C_\mu 
\nonumber \\
&= 2  \pi_\al^\rho \pi_\be^\mu g^{\la\si} \partial_{\la\si\rho }  C_\mu 
 + B_{\al\be}(g, \partial g, \partial^2 g, \partial^3 g, C, \partial C, \partial^2 C).
\nonumber
\end{align}
But
\begin{gather}
\nabla_\la \nabla_\si \nabla_\rho C_\mu = \partial_{\la\si\rho} C_\mu 
+ B_{\la\si\rho \mu}(g, \partial g, \partial^2 g, \partial^3 g, C, \partial C, \partial^2 C),
\nonumber
\end{gather}
and commuting the covariant derivatives gives
\begin{gather}
\nabla_\la \nabla_\si \nabla_\rho C_\mu
= 
\nabla_\rho \nabla_\la \nabla_\si  C_\mu + 
B_{\la\si\rho \mu}(g, \partial g, \partial^2 g, \partial^3 g, C, \partial C ),
\nonumber
\end{gather}
so that
\begin{gather}
 \partial_{\la\si\rho} C_\mu  =\nabla_\rho \nabla_\la \nabla_\si  C_\mu 
+ B_{\la\si\rho \mu}(g, \partial g, \partial^2 g, \partial^3 g, C, \partial C, \partial^2 C).
\nonumber
\end{gather}
Contracting with $g^{\la\si}$ and invoking (\ref{Laplacian_curvature}) 
produces
\begin{align}
\begin{split}
g^{\la\si}\partial_{\la\si\rho} C_\mu  
& =
\nabla_\rho  ( -\Delta C_\mu + R_\mu^\tau C_\tau ) 
+ B_{\rho \mu}(g, \partial g, \partial^2 g, \partial^3 g, C, \partial C, \partial^2 C)
\\
& =B_{\rho \mu}(g, \partial g, \partial^2 g, \partial^3 g, C, \partial C, \partial^2 C),
\end{split}
\nonumber
\end{align}
since  $\Delta C = 0$ by (\ref{Hodge_Laplacian}),
(\ref{incompressible}), and 
(\ref{irrotational}).

Apply  $g^{\rho \mu}\partial_{\rho \mu}$ to (\ref{entropy_eq}) 
and use $\frac{r\theta}{F} > 0$ to find
\begin{gather}
g^{\rho\mu} C^\al \partial_{\rho\mu\al} s =
B( g, \partial g, \partial^2 g, \partial^3 g, s, \partial s, \partial^2 s, C, \partial C, 
\partial^2 C),
\label{entropy_eq_simple}
\end{gather}
where third derivatives of $C$ are not present by an argument similar to the one
used above involving $\Delta C$ in the
derivation of (\ref{Einstein_simple_fourth_order}).

Finally, apply $g^{\la\mu}\partial_{\la\mu}$ to (\ref{current_eq_simple_second_order}) 
and use   (\ref{irrotational}) to get
\begin{align}
\begin{split}
g^{\rho\mu}g^{\la\si} 
\partial_{\rho\mu\la \si} C_\al = 
 B_\al(g, \partial g, \partial^2 g, \partial^3 g, \partial^4 g, 
 C, \partial C, \partial^2 C, \partial^3 C).
\end{split}
\label{current_eq_simple_no_Omega}
\end{align}

As we shall see, the system of 15 equations 
(\ref{Einstein_simple_fourth_order}), (\ref{entropy_eq_simple}),
and (\ref{current_eq_simple_no_Omega}), 
for the ten $g_{\al\be}$, one $s$, and four $C_\al$ (again, $F$
is considered as a function of $C^\al$ and $g_{\al\be}$) forms a Leray-Ohya
system with index structure (see the appendix for definitions)
 \begin{gather}
\begin{cases}
 \mathbf{m}(\text{eq. } (\ref{Einstein_simple_fourth_order}) ) = 4, \,
\mathbf{m}(\text{eq. } (\ref{entropy_eq_simple}) ) = 3, \,
\mathbf{m}(\text{eq. } (\ref{current_eq_simple_no_Omega} )) = 4, \,
\\
\mathbf{s}(g_{\al\be}) = 5, \, \mathbf{s}(s) = 4, \, 
\mathbf{s}(C_\al) = 4, 
\\
 \mathbf{t}(\text{eq. } (\ref{Einstein_simple_fourth_order}) ) = 2, \,
\mathbf{t}(\text{eq. } (\ref{entropy_eq_simple}) ) = 2, \,
\mathbf{t}(\text{eq. } (\ref{current_eq_simple_no_Omega} )) = 1.
\end{cases}
\label{index_structure}
 \end{gather}
In order to confirm this and apply theorem \ref{Leray-Ohya_theorem} from 
the appendix,
we need to first turn our attention to the Cauchy data and the system's
characteristics. 

\section{Initial data and characteristics.\label{initial_data_section}}

\subsection{Characteristics.\label{characteristics_section}}
Consider a regular hypersurface $\cS$  that is locally given as the zero
 set of a sufficiently differentiable function, i.e., $\cS = \{ f(x) = 0 \}$.
Looking at the left hand side of the system
(\ref{current_eq_simple_second_order}), (\ref{Helmholtz_simple}), (\ref{Einstein_simple}),
(\ref{entropy_eq_simple_first_order}), we see at once that 
$\cS$ will be characteristic if any of the two following conditions hold:
\begin{gather}
C^\al \partial_\al f = 0,
\label{flow_lines_characteristic}
\end{gather}
or
\begin{gather}
g^{\al\be} \partial_\al f \partial_\be f = 0.
\label{light_cone_characteristic}
\end{gather}
A quick inspection shows that these are also the characteristics of the 
system
(\ref{Einstein_simple_fourth_order}), (\ref{entropy_eq_simple}),
and (\ref{current_eq_simple_no_Omega}), although  
the characteristics of 
(\ref{Einstein_simple_fourth_order}) and (\ref{current_eq_simple_no_Omega})
have multiplicity two.

Recalling that $C^\al = F u^\al$, we see that (\ref{flow_lines_characteristic})
expresses that $\cS$ is spanned by the flow-lines of the fluid,
whereas (\ref{light_cone_characteristic}) means that $\cS$ is tangent to the 
light-cone of the metric $g$ at each point. The reader should
 contrast the present characteristic surfaces with those of 
 a relativistic inviscid fluid. In the latter case, besides the flow-lines and 
surfaces tangent to the light-cone, a third family of physically meaningful
characteristic surfaces is present, namely, those corresponding to the sound
waves of the fluid. As it is discussed in section \ref{comparison_section},
if no simplifying assumption is made, and we consider
the full set of equations (\ref{entropy_eq}), 
 (\ref{Helmholtz}),  (\ref{current_eq_original}), and 
 (\ref{Einstein_simple}),
a third family of characteristics also appears in the viscous case, 
but these are non-physical and do not correspond
to the propagation of any physical quantity. In fact,  if one insists in computing the speed of propagation of would-be acoustic waves along such surfaces,
 it is found to be infinite. Hence, here, 
as in the non-relativistic case, there is not a well-defined
notion of sound speed for a viscous fluid.

\subsection{Cauchy data for an incompressible fluid.\label{Cauchy_incompressible_section}}
We are interested in obtaining initial conditions for 
the system (\ref{Einstein_simple_fourth_order}), (\ref{entropy_eq_simple}),
and (\ref{current_eq_simple_no_Omega}). 
Initially, we do not necessarily assume
 (\ref{irrotational}), 
and along the way, initial data for 
equations  (\ref{current_eq_simple_second_order}), (\ref{Helmholtz_simple}), (\ref{Einstein_simple}),
(\ref{entropy_eq_simple_first_order}) will be obtained.
 Hence, we suppose we are given a solution
to the incompressible Einstein-Navier-Stokes equations. As 
in section \ref{new_system_section}, we continue
to assume
that $s > 0$ and $F > 0$ are the two independent thermodynamic variables,
with an equation of state $r =r(F, s)$
satisfying the hypotheses stated in Theorem \ref{main_theorem}.

The Cauchy data\footnote{We remark that we shall eventually consider 
harmonic coordinates, in which case 
only the spatial components $g_{ij}$,
$\partial_0 g_{ij}$ and $u^i$  on $\Si$ are given in the initial Cauchy data. The
remaining components are obtained by the use of harmonic coordinates
and (\ref{original_normalization}). See section \ref{proof_secion}.} that is given for system 
(\ref{original}) consist of the values of
\begin{gather}
 g_{\al\be},  \partial_0 g_{\al\be}, u_\al, s, \text{ and } F \text{ on } \Si.
\label{Cauchy_data}
\end{gather}
We naturally suppose that the constraints are satisfied and that 
 $\Si$ is non-characteristic. As a consequence,
in the principal part of system
(\ref{current_eq_simple_second_order}), (\ref{Helmholtz_simple}), (\ref{Einstein_simple}),
(\ref{entropy_eq_simple_first_order}), and of system
 (\ref{Einstein_simple_fourth_order}), (\ref{entropy_eq_simple}),
and (\ref{current_eq_simple_no_Omega}), one can always solve for the highest
time derivatives appearing on the left-hand side in terms of quantities on the 
right-hand side, and this fact will be extensively used below;
derivatives along $\Si$ can always be computed and present no problem
in determining the Cauchy data.

We shall say that some expression 
is equal to $C.D.$ (Cauchy data) when it can be expressed solely in terms of
quantities that are written in terms of the Cauchy data on $\Si$. 
$\overline{\partial}$ will symbolically denote spatial derivatives $\partial_i$. 
We shall denote by $Z$ a general
smooth function of its arguments, which may vary from expression to expression.

Recall  that 
\begin{gather}
C^\al = F u^\alpha,
\label{C_F_U} 
\end{gather}
so in particular $C^\al$ is known on $\Si$.

We derive the values of $\partial_0 u_\al$ on $\Si$. Differentiating
(\ref{original_normalization}) yields $u^\al \nabla_0 u_\al = 0$, so that
\begin{gather}
u^\al \partial_0 u_\al = Z(g, \partial g, u, \overline{\partial} u ).
\nonumber
\end{gather}
Pichon \cite{Pichon} has shown that if $u$ is time-like, then the above relation 
used in conjunction with the constraints 
$S_\al^0 = \K T_\al^0$ and (\ref{original_normalization}) allow
us to solve for the values of $\partial_0 u_\al$ by first changing to a coordinate
system where $u^i = 0$. We have therefore 
determined $\partial_0 u_\al$ on $\Si$, i.e., 
\begin{gather}
\left. \partial_0 u_\al \right|_\Si  = \left. Z(g,\partial g, u, \overline{\partial} u) \right|_\Si = C.D. 
\label{partial_0_u_CD}
\end{gather}
Using (\ref{C_F_U}) in (\ref{entropy_eq})
yields
\begin{gather}
r \theta u^\al \partial_\al s - F \sv (
\pi^{\al\rho} \nabla_\rho u_\be \nabla_\al u^\be +
 \nabla_\al u^\mu \nabla_\mu u^\alpha) = 0.
\nonumber
\end{gather}
Using (\ref{partial_0_u_CD}), $r, \theta > 0$,
\begin{gather}
\left. \partial_0 s \right|_\Si =  C. D. 
\label{partial_0_s_CD}
\end{gather}
Notice that (\ref{partial_0_s_CD})  could not have been determined directly from
(\ref{entropy_eq_simple_first_order}) in that this last equation involves
 $\partial_\mu C_\al$ which, as we next show, requires 
 $\left. \partial_\mu s \right|_\Si $ to be known on $\Si$.

From (\ref{original_mass_conservation}) we obtain
\begin{gather}
u^\al \frac{\partial r}{\partial s} \partial_\al s +  \frac{\partial r}{\partial F} u^\al 
\partial_\al F + r \nabla_\al u^\al = 0.
\nonumber
\end{gather}
Since $\frac{\partial r}{\partial F} > 0$ 
by (\ref{sound_speed_condition}) and $\partial_0 s $ is known on $\Si$
by (\ref{partial_0_u_CD}),  we also have
\begin{gather}
\left. \partial_0 F\right|_\Si  = C. D.
\label{partial_0_F_CD}
\end{gather}
It follows from (\ref{partial_0_u_CD}) and (\ref{partial_0_F_CD}) that
\begin{gather}
\left. \partial_0 C_\al \right|_\Si = C. D. 
\label{partial_0_C_CD}
\end{gather}
From (\ref{Einstein_simple}) and the above we get
\begin{gather}
\left. \partial_{00} g_{\al\be} \right|_\Si = C.D.
\nonumber
\end{gather}

From $\pi^{\ga \mu} \nabla^\al T_{\al\mu} =  0$ we obtain
\begin{align}
\begin{split}
\sv \pi^{\ga \mu} \pi^{\rho\al}
(\partial_{\al \rho}C_\mu + \partial_{\al\mu} C_\rho)
= Z(g, \partial g, \partial^2 g, F, \partial F, s, \partial s, C, \partial C, \partial  C).
\end{split}
\label{to_determine_partial_2_C}
\end{align}
We shall use a similar procedure as  employed by Pichon \cite{Pichon}.
We can choose coordinates near $\Si$ such that 
\begin{gather}
u^0 \neq 0, u^i = 0, \pi^\mu_0 = 0, \pi^{00} \neq 0,
\label{special_coord_u}
\end{gather}
so that (\ref{to_determine_partial_2_C})
becomes, after lowering the index $\ga$,
\begin{align}
\begin{split}
\sv
(\pi^{00} \pi^\mu_\ga + \pi^0_\ga \pi^{\mu0} ) \partial_{00} C_\mu 
= Z(g, \partial g, \partial^2 g, F, \partial F, s, \partial s, C, \partial C, 
\overline{\partial} \partial C).
\end{split}
\label{algebraic_system_delta_C}
\end{align}
Setting $\ga = 0$ in (\ref{algebraic_system_delta_C}) and
 using (\ref{original_normalization}) yields
\begin{align}
\begin{split}
\sv
(u^0)^2 \partial_{00} C_0
= Z(g, \partial g, \partial^2 g, F, \partial F, s, \partial s, C, \partial C, 
\overline{\partial} \partial C).
\end{split}
\nonumber
\end{align}
Putting $\ga = i$ in (\ref{algebraic_system_delta_C}):
\begin{gather}
\sv (\de^j_i \pi^{00} + \pi_i^0 \pi^{j0} ) \partial_{00} C_j = 
 Z(g, \partial g, \partial^2 g, F, \partial F, s, \partial s, C, \partial C, 
\overline{\partial} \partial C).
\nonumber
\end{gather}
The determinant of the matrix on the left-hand side is 
\begin{gather}
2 \sv^3 (\pi^{00})^3 \neq 0,
\nonumber
\end{gather}
(where we used that $\pi^0_\mu \pi^{\mu 0} = \pi^0_i \pi^{i0} = \pi^{00}$)
 implying that $\left. \partial_{00} C_\mu \right|_\Si$ is known 
since, in light of (\ref{partial_0_C_CD}),
$\left. \partial_0 C_\mu \right|_\Si$  and the other terms on the right 
hand side are  known on $\Si$ as shown above.
For a general coordinate system,
one changes coordinates to satisfy (\ref{special_coord_u}),
where all second derivatives of $C$ on $\Si$ can be found.
Changing coordinates back gives that
\begin{gather}
\left. \partial_{00} C_\mu \right|_\Si = C.D.
\label{partial_00_C_CD}
\end{gather}
Notice that we could not have determined 
$\left. \partial_{00} C_\mu \right|_\Si $ from (\ref{current_eq_simple_second_order})
since this involves first derivatives of $\Om$.
In addition to (\ref{Cauchy_data}), we have therefore determined
\begin{gather}
\partial^2 g, \partial s,  C, \partial C, \text{ and } \partial^2 C \text{ on } \Si.
\label{Cauchy_data_1}
\end{gather}
Also, from (\ref{contraction_Omega}), (\ref{L_be}), (\ref{V_be}),
 (\ref{Cauchy_data}), and (\ref{Cauchy_data_1}), we see that
 \begin{gather}
 \left. \Om_{\al\be} \right|_\Si = C. D.
 \label{Omega_CD}
 \end{gather}
Differentiating (\ref{entropy_eq_simple_first_order}), using 
(\ref{Cauchy_data}) and (\ref{Cauchy_data_1}) gives
\begin{gather}
\left. \partial_{00} s \right|_\Si = C. D.,
\label{partial_00_s_CD}
\end{gather}
and differentiating (\ref{Einstein_simple}), using 
(\ref{Cauchy_data}),  (\ref{Cauchy_data_1}), and  (\ref{partial_00_s_CD}) 
gives
\begin{gather}
\left. \partial^3_{0} g_{\al \be} \right|_\Si = C. D.,
\label{partial_000_g_CD}
\end{gather}
Differentiating (\ref{to_determine_partial_2_C}), using a similar argument
as the one leading to (\ref{partial_00_C_CD}) and invoking 
(\ref{Cauchy_data}),  (\ref{Cauchy_data_1}),   (\ref{partial_00_s_CD}), 
and (\ref{partial_000_g_CD}) yields
\begin{gather}
\left. \partial^3_0 C_\al \right|_\Si = C. D. 
\label{partial_000_C_CD}
\end{gather}
Differentiating (\ref{contraction_Omega}), using 
(\ref{L_be}), (\ref{V_be}),
 (\ref{Cauchy_data}),  (\ref{Cauchy_data_1}),
 (\ref{Omega_CD}),
 (\ref{partial_00_s_CD}),
 (\ref{partial_000_g_CD}), and (\ref{partial_000_C_CD}),  gives
\begin{gather}
\left. \partial_0 \Om_{\al\be} \right|_\Si = C. D. 
\label{partial_0_Omega_CD}
\end{gather}
Finally, take two derivatives of (\ref{Einstein_simple}),
use (\ref{Cauchy_data}), (\ref{Cauchy_data_1}), (\ref{partial_00_s_CD}), (\ref{partial_000_g_CD}), (\ref{partial_000_C_CD}) to get
\begin{gather}
\left. \partial^4_0 g_{\al\be} \right|_\Si = C. D. 
\label{partial_0000_g_CD}
\end{gather}
Together, (\ref{Cauchy_data}), and (\ref{Cauchy_data_1}),
 (\ref{Omega_CD}), (\ref{partial_00_s_CD}),
 (\ref{partial_000_g_CD}), (\ref{partial_000_C_CD}),  
 (\ref{partial_0_Omega_CD}), and (\ref{partial_0000_g_CD}) are the desired Cauchy data
 for the system (\ref{current_eq_simple_second_order}), (\ref{Helmholtz_simple}),
 (\ref{Einstein_simple}), and 
(\ref{entropy_eq_simple_first_order}). It is clear 
 that we can continue the above process, and determine
 all the derivatives of the unknowns on $\Si$. If the data is
 analytic, we obtain an analytic solution in a neighborhood of $\Si$.
 
 \subsection{Cauchy data for an  incompressible irrotational fluid.
 \label{Cauchy_irroatational_section}}
 We now turn our attention to system
  (\ref{Einstein_simple_fourth_order}), (\ref{entropy_eq_simple}),
and (\ref{current_eq_simple_no_Omega}), so we suppose from now on that
$\Om = 0$. In view of
(\ref{index_structure}) and Theorem \ref{Leray-Ohya_theorem}, 
we need to determine the derivatives of $g_{\al\be}$ up to order $4$; of 
$s$ up to order $3$;  and of $C_\al$ up to order $3$. 

Noticing that the initial data derived in section \ref{Cauchy_incompressible_section} is 
compatible with equations (\ref{Einstein_simple_fourth_order}),
 (\ref{entropy_eq_simple}), and (\ref{current_eq_simple_no_Omega}),
 it remains to find the third derivatives of $s$ on $\Si$.
Using  (\ref{entropy_eq_simple}),
(\ref{Cauchy_data}), (\ref{Cauchy_data_1}), (\ref{partial_00_s_CD}),  (\ref{partial_000_g_CD}), and (\ref{partial_000_C_CD}) gives
\begin{gather}
\left. \partial_0^3 s \right|_\Si = C. D.
\label{partial_000_s_CD}
\end{gather}
(\ref{Cauchy_data}), (\ref{Cauchy_data_1}), (\ref{partial_00_s_CD}),  (\ref{partial_000_g_CD}),  (\ref{partial_000_C_CD}), (\ref{partial_0000_g_CD}),
and (\ref{partial_000_s_CD}) are the initial data for 
equations  (\ref{Einstein_simple_fourth_order}), (\ref{entropy_eq_simple}),
and (\ref{current_eq_simple_no_Omega}). These have to satisfy further
compatibility conditions, as explained in appendix \ref{Leray_Ohya}.
These are determined by plugging the initial data into equations 
 (\ref{Einstein_simple_fourth_order}), (\ref{entropy_eq_simple}),
and (\ref{current_eq_simple_no_Omega}), and taking 
$\mathbf{t}(\text{eq. } (\ref{Einstein_simple_fourth_order}) )  - 1=2- 1 = 1$
derivative of equation (\ref{Einstein_simple_fourth_order})
and $\mathbf{t}(\text{eq. } (\ref{entropy_eq_simple}) ) - 1 = 2- 1 = 1$
derivative of equation (\ref{entropy_eq_simple});
no further derivative of equation (\ref{current_eq_simple_no_Omega}) 
is necessary since $\mathbf{t}(\text{eq. } (\ref{current_eq_simple_no_Omega} )) 
-1 = 1 - 1 = 0$.

\section{Proof of Theorem \ref{main_theorem}.\label{proof_secion}}

We shall show that under the hypotheses stated in Theorem 
\ref{main_theorem}, the system 
(\ref{Einstein_simple_fourth_order}), (\ref{entropy_eq_simple}),
and (\ref{current_eq_simple_no_Omega})
is a Leray system, and Theorem \ref{Leray-Ohya_theorem} can be applied.
The reader is referred to appendix \ref{Leray_Ohya} for the notation and 
terminology employed in this section in connection with Theorem 
\ref{main_theorem}.

Let $\cI$ be given as in the statement of Theorem \ref{main_theorem}. 
Following a standard procedure, we embed $\Si$ into 
the product $\RR \times \Si$,  and  consider a coordinate
chart $\cU \subset \Si$, where harmonic coordinates have been chosen
 such that on $\cU$, $g_{ij} = (g_{0})_{ij}$,  $\partial_0 g_{ij} = \kappa_{ij}$, 
$g_{00} = 1$, $g_{0i} = 0$, with $\partial_0 g_{\al0}$  determined by 
the conditions of harmonic coordinates on $\cU$. On $\cU$ we also have
$u^i = v^i$ and $u^0$ determined by the condition (\ref{original_normalization}).
The values of $s$ and $r$ are known on $\cU$ from 
the initial data, which determines, via the equation of state, 
the values of $F$ on $\cU$. We have, therefore, 
the Cauchy data (\ref{Cauchy_data}).
As shown in section \ref{Cauchy_irroatational_section}, from this
the Cauchy data for the system 
(\ref{Einstein_simple_fourth_order}), (\ref{entropy_eq_simple}),
and (\ref{current_eq_simple_no_Omega}), is known.

Let $A$ be the principal part of our system, i.e., the matrix formed
by the left-hand side of equations 
(\ref{Einstein_simple_fourth_order}), (\ref{entropy_eq_simple}),
and (\ref{current_eq_simple_no_Omega}). Symbolically:
\begin{gather}
A(x,g, s, C, \partial) = 
\left(
\begin{matrix}
g^{\la\si} g^{\mu\rho} \partial_{\la\si\mu\rho} & 0 & 0 \\
0 & g^{\rho \mu} C^\al \partial_{\rho \mu \al} & 0 \\
0 & 0 & g^{\la\si} g^{\mu\rho} \partial_{\la\si\mu\rho}
\end{matrix}
\right).
\nonumber
\end{gather}
Let $a_1 = a_3 =  g^{\la\si} g^{\mu\rho} \partial_{\la\si\mu\rho}$, 
$a_2=  g^{\rho \mu} C^\al \partial_{\rho \mu \al}$, 
and let $h_1(x,\xi) = h_3(x,\xi)$,  and $h_2(x,\xi)$ be their characteristic
polynomials when $g_{\al\be}$, $s$, and $C^\al$ are given. 
To identify the characteristic cones,
set 
\begin{gather}
h_2(x,\xi) = g^{\rho \mu}(x) C^\al(x) \xi_\rho \xi_\mu \xi_\al = 0,
\nonumber
\end{gather}
for $x \in \cU$,  with $g^{\rho \mu}$ and $C^\al$ replaced by the 
corresponding Cauchy data on $\cU$. It follows that $h_2(x,\xi)$ is
 hyperbolic at $x$ if  $g_{\al\be}(x) C^\al(x) C^\be(x) > 0$, i.e., if 
 $C$ is time-like with respect to the metric $g$. Since $C^\al = F u^\al$, 
 this is the case by the hypotheses of Theorem \ref{main_theorem}. 
 A consequence is that the half-cone defined by $C^\al \xi_\al \geq 0$ is 
 exterior to the one given by $g^{\al\be} \xi_\al\xi_\be \geq 0$, hence
 $\Ga_x^+(a_2)$ coincides with the half-cone 
  $g^{\al\be} \xi_\al\xi_\be \geq 0$ of the metric $g$. 
For $a_1$ and $a_3$ we have
\begin{gather}
h_1(x,\xi) = g^{\rho \mu}(x) g^{\la\si}(x) \xi_\rho \xi_\mu \xi_\la \xi_\si
= ( g^{\rho \mu}(x) \xi_\rho \xi_\mu )^2 = 0,
\nonumber
\end{gather}
i.e., the characteristic polynomial $h_1(x,\si)$ is the product of the two
hyperbolic polynomials given by the metric $g$. 
We conclude that the operator
$A(x,g, s, C, \partial)$ is Leray-Ohya hyperbolic for any $x \in \cU$ with
half-cones $\Ga^\pm_x(A)$ agreeing with those of the metric $g$, and
$p_1 = p_3 = 2$, $p_2 = 1$. Thus $w = 2$, and  the initial data
belongs to Gevrey spaces satisfying the condition $1 \leq  \si < 2 = \frac{w}{w-1}$.

Next, write the system as
\begin{gather}
A(x,g, s, C, \partial)(g,s,C) = B(x, g, s, C),
\label{system_short}
\end{gather}
where $B$ is given by the expression on the right-hand side 
of equations  (\ref{Einstein_simple_fourth_order}), (\ref{entropy_eq_simple}),
and (\ref{current_eq_simple_no_Omega}). 
The matrix $A$ depends polynomially on the functions
 $g$, $s$, and $C$, while $B$ is a rational function
of these variables and their derivatives. 
The denominator of the rational expressions
appearing in $B$ are of the form $F = \sqrt{C^\al C_\al}$,
 $F^2 = C^\al C_\al$,
$r=r(F,s)$, $\theta=\theta(F,s)$ or products of them.
By our hypotheses, all such terms are uniformly bounded away from 
zero on $\Si$. And in view of our choice of indices (\ref{index_structure}), 
it follows that the coefficients of system
(\ref{system_short}) satisfy the hypotheses of Theorem \ref{Leray-Ohya_theorem}
with $m = 3$.

It remains to verify that (\ref{system_short}) is indeed a Leray system
with  the index structure (\ref{index_structure}). The orders of the operators
are 
$\mathbf{m}(\text{eq. } (\ref{Einstein_simple_fourth_order}) ) = 4 
=  \mathbf{s}(g_{\al\be})  -  \mathbf{t}(\text{eq. } (\ref{Einstein_simple_fourth_order}) )
+ 1 =$ order of $a_1$;
$\mathbf{m}(\text{eq. } (\ref{entropy_eq_simple}) ) = 3
= \mathbf{s}(s)  - 
\mathbf{t}(\text{eq. } (\ref{entropy_eq_simple}) ) + 1 = $ order of $a_2$;
$\mathbf{m}(\text{eq. } (\ref{current_eq_simple_no_Omega} )) = 4
= \mathbf{s}(C_\al) - \mathbf{t}(\text{eq. } (\ref{current_eq_simple_no_Omega} ))+1
=$ order of $a_3$. The coefficients of $a_j$, $j=1,2,3$, 
do not depend on derivatives of $g$, $s$, or $C$, so it suffices
to verify that each $b_t(x,g,s,C)$ depends
on at most $\bs(k) - \bt(t)$ derivatives of the corresponding $k^{\text{th}}$
unknown. Below we list, for each equation, the difference 
$\bs(k) - \bt(t)$ and the corresponding highest order derivative appearing
on the right-hand side of the equation.
\begin{gather}
\text{eq. } (\ref{Einstein_simple_fourth_order})
\begin{cases}
\mathbf{s}(g_{\al\be}) -  \mathbf{t}(\text{eq. } (\ref{Einstein_simple_fourth_order}) ) 
= 3, & \partial^3g, \\
\mathbf{s}(s) -  \mathbf{t}(\text{eq. } (\ref{Einstein_simple_fourth_order}) ) 
= 2, & \partial^2s, \\
\mathbf{s}(C_\al) -  \mathbf{t}(\text{eq. } (\ref{Einstein_simple_fourth_order}) ) 
= 2, & \partial^2 C,
\end{cases}
\nonumber
\end{gather}
\begin{gather}
\text{eq. } (\ref{entropy_eq_simple})  
\begin{cases}
\mathbf{s}(g_{\al\be}) -  \mathbf{t}(\text{eq. } (\ref{entropy_eq_simple}) )  
= 3, & \partial^3g, \\
\mathbf{s}(s) -  \mathbf{t}(\text{eq. } (\ref{entropy_eq_simple}) )   
= 2, & \partial^2s, \\
\mathbf{s}(C_\al) -  \mathbf{t}(\text{eq. } (\ref{entropy_eq_simple}) )   
= 2, & \partial^2 C,
\end{cases}
\nonumber
\end{gather}
\begin{gather}
\text{eq. } (\ref{current_eq_simple_no_Omega} ) 
\begin{cases}
\mathbf{s}(g_{\al\be}) -  \mathbf{t}(\text{eq. } (\ref{current_eq_simple_no_Omega} ) )  
= 4, & \partial^4 g, \\
\mathbf{s}(s) -  \mathbf{t}(\text{eq. }(\ref{current_eq_simple_no_Omega} )) )   
= 3, & \varnothing,  \\
\mathbf{s}(C_\al) -  \mathbf{t}(\text{eq. } (\ref{current_eq_simple_no_Omega} ) )   
= 3, & \partial^3 C,
\end{cases}
\nonumber
\end{gather}
where $\varnothing$ indicates that the corresponding variable does not appear.
We have therefore verified all the hypotheses of Theorem
\ref{Leray-Ohya_theorem}, obtaining in this way a solution to
equations (\ref{Einstein_simple_fourth_order}), (\ref{entropy_eq_simple}),
and (\ref{current_eq_simple_no_Omega})
in a neighborhood of $\cU$. This solution is the unique solution
in the Gevrey class indicated in Theorem \ref{main_theorem}.

We need to show that this solution yields a solution to the original 
incompressible irrotational Einstein-Navier-Stokes system.
Suppose first that the initial data given in the hypotheses
of Theorem \ref{main_theorem} is analytic, and consider the 
Einstein-Navier-Stokes system with the reduced Einstein equations
in place of (\ref{original_Einstein}), i.e., suppose 
that the system is  written in harmonic coordinates.
Pichon \cite{Pichon} has shown how the analytic Cauchy problem
for the Einstein-Navier-Stokes  system, with the reduced Einstein equations
in (\ref{original_Einstein}), can be solved by successively 
computing higher order time derivatives in terms of the Cauchy data
on $\Si$. His work only treated the case of an equation of state
$p = p(\varrho)$, i.e., without including entropy, but it is not 
difficult to see that his procedure can be  generalized to allow
for the more general system we are treating here. 
By the way the Cauchy data  
was derived in section \ref{Cauchy_incompressible_section},
and upon setting
$C^\al = F u^\al$ and $\Om_{\al\be} = 
\nabla_\al C_\al - \nabla_\be C_\be$,
this solution satisfies equations
 (\ref{current_eq_simple_second_order}), (\ref{Helmholtz_simple}), (\ref{Einstein_simple}), and
(\ref{entropy_eq_simple_first_order})
when (\ref{T_vis}) reduces to (\ref{T_vis_i}), i.e., 
when the fluid is incompressible. If, furthermore, the fluid
is also irrotational, i.e., (\ref{T_vis_i}) reduces to (\ref{T_vis_ii}),
then this solution also satisfies 
(\ref{Einstein_simple_fourth_order}), (\ref{entropy_eq_simple}), and (\ref{current_eq_simple_no_Omega}) in light of the
way its initial data was obtained in section
\ref{Cauchy_irroatational_section}, and it necessarily agrees
with the solution given by the Leray-Ohya Theorem.

Summing up, if we are given analytic initial data for the 
incompressible irrotational Einstein-Navier-Stokes equations, then 
we obtain an analytic solution to the system (\ref{original}) written 
in harmonic coordinates, and this solution satisfies the system  
(\ref{Einstein_simple_fourth_order}), (\ref{entropy_eq_simple}), and  (\ref{current_eq_simple_no_Omega}), and coincides with the solution 
given by Leray-Ohya's Theorem.

Returning now to the general, i.e., non-analytic case, we approximate,
in the Gevrey topology,
the given initial data $\cI$ from Theorem \ref{main_theorem}
by a sequence of analytic initial data $\{ \cI_\nu \}$, 
obtaining a family of 
analytic solutions $\{ Z_\nu = (g_\nu, u_\nu, s_\nu, r_\nu) \}$ 
to the incompressible irrotational Einstein-Navier-Stokes system
written in harmonic coordinates. This  yields a family of solutions
 $\{ U_\nu = (g_\nu, s_\nu, C_\nu) \}$ to system
(\ref{Einstein_simple_fourth_order}), (\ref{entropy_eq_simple}), and (\ref{current_eq_simple_no_Omega}). When $\cI_\nu \rar \cI$,
one gets 
$U_\nu \rar U$, where $U$ is the solution obtained above by application 
of the Leray-Ohya Theorem. The energy-type of estimates derived by Leray 
and Ohya \cite{Leray_Ohya} guarantee that the sequence 
$Z_\nu$ also has a limit $Z$ that lies in the stated Gevrey spaces and satisfies
the system (\ref{original}) written in harmonic coordinates, with the
stress-energy tensor given by (\ref{T_vis_ii})\footnote{We notice that this
approximation argument is essentially the same one used by Lichnerowicz
to produce solutions in the Gevrey class for the equations of
 magneto-hydrodynamics \cite{Lich_MHD}.}. It is well-known that
a solution to the reduced Einstein equations satisfies the full system
if and only if the constraints are satisfied, which is the case by our
hypotheses on the initial data. We also notice that from 
$\pi^{\ga \be} \nabla^\al T_{\al \be}=0$
it follows that 
\begin{gather}
0 = u^\al u^\ga \nabla_\al u_\ga = \frac{1}{2} u^\al\partial_\al (u^\ga u_\ga),
\nonumber
\end{gather}
and therefore $u$, being unitary at time zero, remains unitary.

The existence of a domain of  dependence, as stated 
in Theorem \ref{main_theorem},
follows at once 
from the domain of dependence property of  
Theorem \ref{Leray-Ohya_theorem}, using the fact (shown above) that 
the half-cones $\Ga^\pm_x(A)$ agree with those of the metric $g$.
With the domain of dependence property at hand, a standard gluing argument
now gives a global, in space, solution that is geometrically unique.
This finishes the proof of Theorem \ref{main_theorem}.

\begin{remark} An natural question is whether the condition
$\Om = 0$ is preserved under the time-evolution, i.e., whether
$\Om$ remains zero if it vanishes on $\Si$. This is known
to be the case for the non-relativistic Navier-Stokes equations 
\cite{MB}, and
also for 
relativistic inviscid  fluids \cite{Lich_fluid_1}. In order to establish this here,
reason as follows. As mentioned in section \ref{Cauchy_incompressible_section},
the incompressible, not  necessarily irrotational,  system
can be solved for analytic data; notice that such a solution is not
necessarily derived from Theorem \ref{main_theorem}.
In this case, the vorticity satisfies
(\ref{Helmholtz_simple}), and this equations remains valid
when  $\Om = 0$, and in particular when we consider
(\ref{Helmholtz_simple}) with the (analytic) solution given by
Theorem \ref{main_theorem}. On the other hand, 
the analytic solution to (\ref{Helmholtz_simple}),
with initial data given as in Theorem \ref{main_theorem}
and $\Om = 0$ at $t=0$, agrees, by uniqueness, with that given by
the Theorem. Thus $\Om = 0$, if it is zero initially. 
\end{remark}

\begin{remark}
The argument of the previous remark is rather indirect. It is of importance
to obtain a simple evolution equation for $\Om$ only, as in the case
of non-relativistic \cite{MB} and relativistic inviscid fluids
\cite{Lich_fluid_1}. Not only
would this give a direct argument for the propagation of the zero vorticity
condition, but perhaps more importantly, it would give us
a direct way of studying the evolution of the vorticity, a problem 
of clear physical importance.
\end{remark}

\section{Comparison with the inviscid case.\label{comparison_section}}
In this section we derive a different system of equations suited
for comparisons with the case of an ideal fluid. This follows
the spirit of our guiding principle (i) (see introduction), and it is
also a tentative first step in addressing the (notably hard) problem
of the convergence of solutions of the viscous equations to solutions
of the inviscid ones when $\bv, \sv \rar 0$.

We once again consider the Laplacian of the current $C$, i.e.,
equation (\ref{Hodge_Laplacian}), but now a different route will be taken.
Using 
$\de C = - \nabla_\al C^\al = -\nabla_\al(F u^\al)$ and 
$\nabla_\al( r u^\al) = 0$, we get
\begin{gather}
\de C = C^\al ( \frac{ \partial_\al r}{r} - \frac{\partial_\al F }{F} ).
\nonumber
\end{gather}
Set $\widehat{F} = F^2$ and $\cF(\widehat{F}, s) = \log\frac{r}{F}$
(recall that $r = r(F,s)$). So
\begin{gather}
\de C = \iota_C d\cF,
\nonumber
\end{gather}
from which follows, upon computing $d\de C$, that
\begin{gather}
\Delta C = \cL_C d\cF + \de \Om.
\label{Laplacian_C_Lie}
\end{gather}
We recall that $\cL$ is the Lie derivative. Next, we  
compute $\cL_C d\cF$. Computing $\partial_\al \cF$ directly, 
using the definition of $\Om$, and equation 
(\ref{contraction_Omega}) yields
\begin{align}
\partial_\al \cF & = 
2 \frac{\partial \cF}{\partial \widehat{F} } C^\rho \nabla_\rho C_\alpha
+ (\frac{\partial \cF}{\partial s} + 2 \theta F \frac{\partial \cF}{\partial \widehat{F} } )
\partial_\al s \nonumber \\
&  - 2 \frac{\theta}{F} \frac{\partial \cF}{\partial \widehat{F} } C_\al C^\rho
\partial_\rho s 
+ 2 \frac{F}{r} \frac{\partial \cF}{\partial \widehat{F} } (\bv L_\al + \sv V_\al ).
\nonumber
\end{align}
From this and the standard formula $\cL_X w_\al = X^\mu \nabla_\mu w_\alpha+ w_\mu \nabla_\al X^\mu$ we find
\begin{align}
\begin{split}
[\cL_C & d\cF]_\al  = 2 \frac{ \partial \cF}{\partial \widehat{F} } C^\mu C^\rho 
\nabla_\mu \nabla_\rho C_\al + 
(\frac{\partial \cF}{\partial s} + 2 \theta F \frac{\partial \cF }{\partial \widehat{F} } )
(\nabla_\al (C^\mu \nabla_\mu s) - \nabla_\al C^\mu \nabla_\mu s)
 \\
& + C^\mu\nabla_\mu (\frac{\partial \cF}{\partial s} 
+ 2\theta F \frac{\partial \cF}{\partial \widehat{F} } ) \partial_\al s + 2C^\mu \nabla_\mu(\frac{ \partial \cF}{\partial \widehat{F}} C^\rho) \nabla_\rho C_\al 
\\
& 
- 2C^\mu \nabla_\mu ( \frac{\theta}{F} \frac{ \partial \cF }{\partial \widehat{F} }C_\al C^\rho \partial_\rho s) 
+ 2 C^\mu \nabla_\mu \Big(  \frac{F}{r} \frac{\partial \cF}{\partial \widehat{F} } (\bv L_\al + \sv V_\al ) \Big ) + d\cF_\mu \nabla_\al C^\mu.
\end{split}
\nonumber
\end{align}
Combining this with (\ref{Laplacian_C_Lie}), using (\ref{Laplacian_curvature}),
and applying $C^\si \nabla_\si$ to the resulting expression leads to
\begin{align}
\begin{split}
\Big( g^{\mu\rho}& - (1 - \frac{F}{r} \frac{\partial r}{\partial F} ) 
\frac{C^\mu C^\rho}{F^2}  \Big ) C^\si \nabla_\si \nabla_\mu \nabla_\rho C_\al  
+ 
 C^\rho  C^\mu \nabla_\rho \nabla_\mu \Big[ 2 \frac{F}{r} 
\frac{\partial \cF}{\partial \widehat{F} } (\bv L_\al + \sv V_\al ) \Big ]
\\
& - \nabla_\mu(C^\rho \nabla_\rho \Om^\mu_{\mss \al }) 
+ (\frac{\partial \cF}{\partial s } + 2 \theta F \frac{\partial \cF}{\partial \widehat{F} })
C^\si \nabla_\si (C^\mu \nabla_\mu s)  +
\\
& 
 2 C^\si \nabla_\si ( \frac{\partial \cF}{\partial \widehat{F } } C^\mu C^\rho) \nabla_\mu \nabla_\rho C_\al  
+ \nabla_\mu C^\rho \nabla_\rho \Om^\mu_{\mss \al }
+ C^\rho R_{\si\rho} C^\rho \Om^\si_{\mss \al }  +
R_{\al\mss \rho}^{\mss \mu \mss \si} \Om_{\mu \si}
\\&
+  C^\si \nabla_\si (\frac{\partial \cF}{\partial s}+ 2 \theta F \frac{\partial \cF}{\partial \widehat{F } } ) \nabla_\al ( C^\mu \nabla_\mu s)  
  - C^\si \nabla_\si (R_\al^\mu C_\mu)  
  \\
&
+ C^\si \nabla_\si \Big [- (\frac{\partial \cF}{\partial s} + 2\theta F 
\frac{\partial \cF}{\partial \widehat{F} } ) \nabla_\al C^\mu \nabla_\mu s
+ C^\mu \nabla_\mu  (\frac{\partial \cF}{\partial s} + 2\theta F 
\frac{\partial \cF}{\partial \widehat{F} } )  \partial_\al s
\\
& 
 + 2 C^\mu \nabla_\mu ( \frac{\partial \cF}{\partial \widehat{F} } C^\rho ) \nabla_\rho
 C_\al  - 2 C^\mu \nabla_\mu (\frac{\theta}{F} \frac{\partial \cF}{\partial \widehat{F} }
 C_\al C^\rho \partial_\rho s ) \Big ]
 + C^\si \nabla_\si ( d \cF_\mu \nabla_\al C^\mu ) 
 \\
& + C^\rho \nabla_\rho C^\mu \nabla_\mu \Big[ 2 \frac{F}{r} 
\frac{\partial \cF}{\partial \widehat{F} } (\bv L_\al + \sv V_\al ) \Big ]
=0,
\end{split}
\label{current_eq_original_characteristic}
\end{align}
where the Riemann tensor appears due to the commutation of covariant
derivatives.
Taking equations (\ref{entropy_eq}),  (\ref{Helmholtz}),
(\ref{Einstein_simple}), 
and  (\ref{current_eq_original_characteristic}) as the set of equations
for a relativistic 
viscous fluid on the unknowns $s$, $\Om$, $g$, and $C$ (again,
$F$ is treated as a function of $C$ and $g$), we obtain a 
system that has the convenient property of  directly reducing, 
upon  setting 
$\bv = \sv = 0$,
to the system of 
 an inviscid relativistic fluid studied by Lichnerowicz 
 \cite{Lich_fluid_1,Lich_fluid_2}.

Unfortunately, system (\ref{entropy_eq}),  (\ref{Helmholtz}),
(\ref{Einstein_simple}), 
and  (\ref{current_eq_original_characteristic}) does not seem 
amenable to a treatment via the techniques here employed: there are
just too many derivatives present in order to accommodate the necessary
index structure of a Leray system. That is why the
 system  (\ref{entropy_eq}),  (\ref{Helmholtz}),
(\ref{Einstein_simple}), 
and  (\ref{current_eq_original_characteristic}), which would seem 
 natural from the point of view of the equations obtained when 
 $\bv = \sv = 0$, has not been
used, and we had to derive a different set of equations in
order to prove Theorem \ref{main_theorem}. 

 One could try to circumvent the above problem
by incorporating some of the higher order derivatives into the principal part of
the system, i.e., into $A(x,U,\partial )U$. But if one attempts to do this,
the operator $A(x,U,\partial )$ will contain terms of the form
\begin{gather}
(g^{\la\rho} - C^\la C^\rho) \partial_{\la\rho},
\label{bad_characteristic}
\end{gather}
whose characteristic polynomial fails to be hyperbolic. In fact, the
characteristic surfaces defined by 
(\ref{bad_characteristic}) are space-like and correspond to the would-be
acoustic waves mentioned in section \ref{characteristics_section}.
We believe that more specific techniques, tailored to the
special structure of the above equations, will have to be developed
in order to prove, for a general relativistic viscous fluid,
 existence of solutions that enjoy the domain 
of dependence property.

\appendix 

\section{Leray-Ohya systems.\label{Leray_Ohya}}

For the reader's convenience, in this section we briefly 
recall the definition of a Leray-Ohya system and state the Leray-Ohya Theorem  
 used in our proof. 
  
 Let $M$ be a smooth manifold,  $a = a(x,\partial)$, $x \in M$,
a  differential operator  of order $m$ acting on sufficiently regular 
 real valued functions defined on $M$, and $h(x,\partial)$ the principal part
 of $a(x,\partial)$ at $x$.
  For a given $x \in M$, consider the characteristic
 polynomial of $a$ at $x$, denoted by $h(x,\xi)$, where  $\xi \in T_x^* M$; it
  is a homogeneous polynomial of degree $m$. The cone $V_x(h)$ of $h$
 in $T_x^* M$ is defined by the equation
 \begin{gather}
 h(x,\xi) = 0.
 \nonumber
 \end{gather}
$h(x,\xi)$ is called a hyperbolic polynomial if there exists  $\zeta \in T_x^*M$ such 
that every straight line through $\zeta$ that does not contain the origin intersects 
the cone $V_x(h)$ at $m$ real distinct points. Under these
conditions, the set of points $\zeta$ with this  property forms the interior
of two opposite convex half-cones $\Ga_x^{*,+}(a)$, $\Ga_x^{*,-}(a)$, with
$\Ga_x^{*,\pm}(a)$ non-empty, with boundaries that belong to 
$V_x(h)$. 

Consider the $\ell\times \ell$ diagonal matrix
\begin{gather}
A(x,\partial) = 
\left(
\begin{matrix}
a_1(x,\partial) & \cdots & 0 \\
\vdots & \ddots & \vdots \\
0 & \cdots & a_\ell(x,\partial)
\end{matrix}
\right).
\nonumber
\end{gather}
Each $a_t(x,\partial)$, $t = 1, \dots, \ell$ is a differential operator
of order $\bm(t)$. The operator  $A(x,\partial)$ is called Leray-Ohya
 hyperbolic at $x$ 
if:

(i) The characteristic polynomial $h_t(x,\xi)$ of each $a_t(x,\partial)$ is a product 
of  hyperbolic
polynomials, i.e.
\begin{gather}
h_t(x,\xi) = h_t^1(x, \xi) \cdots h_t^{p_t}(x,\xi),\, t=1,\dots, \ell,
\nonumber
\end{gather}
where each $h_t^q(x,\xi)$, $q=1,\dots,p_t$, $t=1,\dots,\ell$,
is a hyperbolic polynomial.

(ii) The two opposite
convex half-cones,
\begin{gather}
\Ga_x^{*,+}(A) = \bigcap_{t=1}^\ell \bigcap_{q=1}^{p_t} \Ga_x^{*,+}(a_t^q),
\, \text{ and } \,
\Ga_x^{*,-}(A) = \bigcap_{t=1}^\ell \bigcap_{q=1}^{p_t} \Ga_x^{*,-}(a_t^q),
\nonumber
\end{gather}
have a non-empty interior. Here, $\Ga_x^{*,\pm}(a_t^q)$ are the
half-cones associated with the hyperbolic polynomials $h_t^q(x,\xi)$,
$q=1,\dots,p_t$,
$t=1,\dots,\ell$.

We define the convex half-cone $C_x^+(A)$ at $T_xM$ as the
set of $v \in T_x M$ such that $\xi(v) \geq 0$ for every
$\xi \in \Ga_x^{*,+}(A)$; $C_x^-(A)$ is analogously defined, and 
we set  $C_x(A) = C_x^+(A) \cup C_x^-(A)$. 
If the convex cones $C_x^+(A)$ and
$C_x^-(A)$ can be continuously distinguished with respect to $x \in M$,
then $M$ is called time-oriented (with respect to the hyperbolic form
provided by the operator $A$). A path in $M$ is called
future (past) time-like with respect to $A$ if its tangent at each point 
$x \in M$ belongs to $C_x^+(A)$ ($C_x^-(A)$); future (past) causal
if its tangent at each point  $x \in M$ belongs or is tangent 
to $C_x^+(A)$ ($C_x^-(A)$). A regular surface $\Si$ is 
called space-like with respect to $A$ if  $T_x\Si$ ($ \subset T_x M$) is
exterior to $C_x(A)$ for each $x \in \Si$.  It follows that for a time-oriented
$M$,
 the concepts of causal past, 
future, domains of dependence and influence of a set can be defined 
in the same way one does when the manifold is endowed with a Lorentzian 
metric. We refer the reader to \cite{Leray_Hyp} for details; here we shall need only 
the following: the causal past $J^-(x)$ of a point $x \in M$ is the set of points that can
be joined to $x$ by a past causal curve. 

Next, we consider the following quasi-linear system of differential 
equations
\begin{gather}
A(x,U,\partial) U = B(x, U),
\nonumber
\end{gather}
where $A(x,U, \partial)$ is the $\ell \times \ell$ diagonal matrix
\begin{gather}
A(x,U, \partial) = 
\left(
\begin{matrix}
a_1(x,U, \partial) & \cdots & 0 \\
\vdots & \ddots & \vdots \\
0 & \cdots & a_\ell(x,U, \partial)
\end{matrix}
\right),
\nonumber
\end{gather}
with $a_t(x,U,\partial)$, $t = 1, \dots, \ell$ differential operators
of order $\bm(t)$, $B(x,U)$ the vector
\begin{gather}
B(x,U) = (b_t(x,U)),\, t=1,\dots, \ell,
\nonumber
\end{gather}
and the vector 
\begin{gather}
U(x) = (u_k(x)), \, k = 1, \dots, \ell
\nonumber
\end{gather}
is the unknown. 

The system $A(x,U,\partial) U = B(x, U)$ is called a Leray system 
if it is possible to attach to each unknown $u_k$ an integer $\bs(k) \geq 1$,
and to each equation $t$ of the system an integer $\bt(t) \geq 1$,
such that\footnote{Notice that the indices $\bs(k)$ and $\bt(t)$ of a Leray system are defined up to an additive constant.}:

(i) $\bm(t) = \bs(t) - \bt(t) + 1$, $t=1,\dots, \ell$;

(ii) the functions $b_t$ and the coefficients of the differential operators
$a_t$ are sufficiently regular functions of $x$, of $u_k$, and of the 
derivatives of $u_k$ of order less than or equal to $\bs(k) - \bt(t)$, 
$k,t=1\dots, \ell$. If for some $k$ and some $t$, $\bs(k) - \bt(t) < 0$, then
the corresponding $a_t$ and $b_t$ do not depend on $u_k$.

A Leray-Ohya system is a Leray system where the matrix $A$ is Leray-Ohya
hyperbolic. In the quasi-linear case, we need to specify the function 
$U$ that  is  plugged into $A(x,U,\partial)$ and then talk about
a Leray-Ohya system for a function $U$. Naturally, the case of interest
is when $U$ assumes the values of the given Cauchy data.

In order to simplify the statements of the Cauchy problem and of the  Leray-Ohya
Theorem, we shall assume from now on that $M$ is diffeomorphic to $\RR^{n+1}$.
This will not affect the applications of these ideas to the Einstein-Navier-Stokes
system because, in light of the domain of dependence or finite propagation
speed property stated in the theorem below,
 it suffices to address the  well-posedness of the system 
from a local point of view.

Let $\Si$ be a regular hypersurface in $M$, which we assume for simplicity to be given by $x^0 = 0$. The Cauchy data on $\Si$ for a Leray 
system in $M$ consists of the values
of $U=(u_k)$ and their derivatives of order less than or equal to $\bs(k) - 1$ on $\Si$,
i.e., $\left. \partial^\al u_k \right|_{\Si}$, $|\al| \leq \bs(k) -1$, $k=1,\dots, \ell$. 
The Cauchy data is required to satisfy the following compatibility conditions. If
$V=(v_k)$ is an extension of the Cauchy data defined in a  neighborhood  
of $\Si$, i.e.
$\left. \partial^\al v_k \right|_{\Si} = \left. \partial^\al u_k \right|_{\Si}$, $|\al| \leq \bs(k) -1$, $k=1,\dots, \ell$, then the difference $a_t(x,V, \partial)V - b_t(x,V)$
and its derivatives of order less than $\bt(t)$ vanish on $\Si$, for
$t=1,\ldots, \ell$. When to a Leray system
 $A(x,U, \partial) U = B(x, U)$
we prescribe initial data satisfying these conditions, we say that we have 
a Cauchy problem for $A(x,U, \partial) U = B(x, U)$. Notice that by definition,
the Cauchy data for a Leray system satisfies the aforementioned compatibility
 conditions.
 
For a number $|X|  > 0$, let $X$ be the strip 
$0 \leq x^0  \leq |X|. $ We 
denote by $\ga_2^{m, (\si)}(X)$ and $\ga_{2, u.l.}^{m,(\si)}(X)$ 
the Gevrey and uniformly local\footnote{The terminology 
uniformly local for Gevrey spaces is not standard and has not been employed 
by Leray and  Ohya.} Gevrey spaces of functions defined on $X$, 
respectively. More precisely, let $S_t = \{ x^0 =t \}$,
\begin{gather}
|D^k u |_t = c(n,k) \sup_{|\al| \leq k} \p D^\al u \p_{L^2(S_t)},
\nonumber
\\
|D^k u |^{u.l.}_t = c(n,k) \sup_{|\al| \leq k, x \in S_t} \p D^\al u \p_{L^2(B^t_1(x) )},
\nonumber
\end{gather}
where $B^t_1(x)$ is the ball of radius one in $S_t$ centered at $x$ and 
$c(n,k)$ is a normalization constant. Then, for $\si \geq 1$,  and
$m$ a non-negative integer,
$ u \in \ga_2^{m, (\si)}(X)$  means that $u \in C^\infty(X)$, and
\begin{gather}
\sup_{|\be| \leq m, \, \al,  \, 0 \leq t \leq |X|} \frac{1}{ (1 + |\al| )^\si } 
\left( |D^{\be + \al} u |_t \right)^\frac{1}{ 1 + |\al|  } < \infty,
 \nonumber
 \end{gather}
where the sup over $\al$ is taken over multi-indices such that $\al_0 = 0$.
Replacing  $ | \cdot  |_t $ by  $ | \cdot  |^{u.l.}_t $ gives $\ga_{2, u. l. }^{m,(\si)}(X)$.
Analogously one defines such spaces when $X$ is an open set of some $R^N$,
and also $\ga_2^{m, (\si)}(X \times Y)$ and $\ga_{2, u.l.}^{m,(\si)}(X \times Y)$. 
See 
\cite{Leray_Ohya} for details. 
 
We can now state Leray-Ohya's result.

\begin{theorem} (Leray-Ohya) On the strip $X \subset \RR^{n+1}$, consider
a Leray system $A(x,U, \partial) U = B(x, U)$, for $U=(u_k)$, $k=1,\ldots, \ell$,
and its Cauchy problem
\begin{gather}
\begin{cases}
A(x,U, \partial) U = B(x, U), & \text{ in } X, \\
D^\al u_k = \varphi^\al , \al| \leq \bs(k) -1, & \text{ on } \Si = \{ x^0 = 0\}.
\end{cases}
\nonumber
\end{gather}
Suppose that $\varphi^\al \in \ga_2^{(\si)}(\Si)$ for some $\si$.
Assume that 
\begin{gather}
a_t(x,y,D) \in \ga_{2,u.l.}^{\bt(t)+m, (\si)}( X \times Y),
\nonumber
\\
b_t(x,y) \in \ga_{2}^{\bt(t)+m, (\si)}( X \times Y),
\nonumber
\end{gather}
where $Y$ is an open set of $R^N$, 
containing the closure of the values of the Cauchy data $\{ \varphi^\al \}$,
 with
$N$ being the number of derivatives of $u_k$ of orders
less than or equal to $\max_k (\bs(k) - \bt(t) )$, 
$t=1, \ldots, \ell$, and $m$ is a non-negative integer.
 Suppose further that the operator $A(x,U,\partial)$
is Leray-Ohya hyperbolic for any $x \in \Si$ when $U$ is replaced by the
Cauchy data, with each characteristic polynomial $h_t(x,\xi)$ a 
product of $p_t$ hyperbolic polynomials, $t=1,\dots, \ell$. 
Set $w = \max_t p_t$. Finally, assume that
$1 \leq \si < \frac{w}{w-1}$, and that 
$m > \frac{n}{2} + \sup_t  p_t - \inf_t \bt(t)$. Then, there exists 
a unique solution $U = (u_k)$ to the above Cauchy problem 
on some strip $X^\prime \subset X$, with
 $u_k \in \ga_2^{\bs(k) + m + 1, (\si)}(X^\prime)$.
 Furthermore, the operator $A(x,U, \partial)$ is Leray-Ohya hyperbolic within
$X^\prime$ (hence the past of points in $ X^\prime$ is defined), 
and the solution enjoys the finite propagation speed property, i.e., 
 for any $x \in X^\prime$, $U(x)$ depends only on
the Cauchy data on $\Si \cap J^-(x)$.
\label{Leray-Ohya_theorem}
\end{theorem} 

We point out that we have not stated the theorem in its most general fashion, 
see \cite{Leray_Ohya}. When $p_t=1$ for all $t$, then the operator $A(x,\partial)$
is called strictly hyperbolic\footnote{Terms like
hyperbolic and strictly hyperbolic have been used in the literature to denote different
concepts. Here we adopt the conventions of Leray and Ohya.}. In this case,
the system is well-posed in Sobolev spaces.

\end{document}